\documentclass[12pt,aps,prb,preprint]{revtex4}   

\usepackage{amsmath}    
\usepackage{graphicx}   
\begin{document}

\title{Generalized Smoluchowski equation with
 correlation between clusters} 
\author{Lionel Sittler}
\affiliation{Theoretische Physik, Fachbereich 8, Universit\"at Wuppertal \\and\\
 Universit\"at Duisburg-Essen, Germany}
\date{\today}
\thanks{I would like to thanks D.E.Wolf for introducing me to the papers ~[\cite{Bales,Pim}]}
\begin{abstract}
In this paper we compute new reaction rates of the Smoluchowski equation 
which takes account of correlations. 
The new rate $K=K^{MF}+K^C$ are the sum of two terms.
The first term is the known Smoluchowski rate with the mean-field 
approximation. The second takes account of a correlation between clusters.
For this purpose we introduce the average path of a 
cluster.
We relate the length of this path to the reaction rate of the Smoluchowski equation.
We solve the implicit dependence between  the average path  and
the density of clusters.
We show that this correlation length is the same for all clusters.
 Our result depends strongly on the spatial dimension $d$.


 The mean-field term $K_{i,j}^{MF}=(D_i+D_j)(r_j+r_i)^{d-2}$, which vanishes for $d=1$
and is valid up to logarithmic correction for $d=2$, is the usual rate found with the Smoluchowski model
 without correlation
(where $r_i$ is the radius and $D_i$ the diffusion constant 
of the cluster).
 We compute  a new rate: the correlation rate
 $K_{i,j}^{C}=(D_i+D_j)(r_j+r_i)^{d-1}M{(\frac{d-1}{d_f}})$ 
 is valid for $d\geq 1$(where $M(\alpha)=\sum_{i=1}^{+\infty}i^{\alpha}N_i$
is the moment of the density of clusters and $d_f$ is the fractal dimension of the cluster).
The result is valid for a large class of diffusion processes and  mass-radius
relations. This approach confirms some analytical solutions in $d=1$ found with 
other methods.
 We show also  Monte-Carlo simulations which illustrate some
exact new  solvable models.
\end{abstract}
\maketitle
There are  different microscopic processes which lead to the 
model with the Smoluchowski equation.
A comprehensive review can be found in ~[\cite{Ernst}]and the 
derivation of the reaction rate can be found in ~[\cite{Fried}]
with the mean-field approach.
 Particles can be injected into the system during time evolution
~[\cite{MBEMF1,MBEMF2,MBEMF3,MBEMF4,Bara}],
the injection of particles can be  periodic in 
time ~[\cite{PLD,Jen}]. For a very anisotropic surface the  model 
is practically  one-dimensional ~[\cite{Pim,Mo}].
Hence we will assume the most general model for an arbitrary space dimension 
$d$ and flux $F(t)$ of particles  injected into the system.
 Our model has two properties:\\
\begin{center}
 \begin{enumerate}
 \item {\it Brownian diffusion}: clusters of mass $k$ diffuse with a Brownian 
motion with a diffusion constant $D_k$.
 \item {\it irreversible aggregation}: clusters interact  through 
a contact process, namely two particles in contact aggregate irreversibly.
 \end{enumerate}\end{center}
The Smoluchowski approach disregards the microscopic density  $n_k(t,\mathbf{r})$
of a cluster of mass $k$
 and considers the macroscopic density, i.e. 
space average density:
$N_k(t)=\frac{1}
{\int d^{(d)}\mathbf{r}}
\int d^{(d)}\mathbf{r} n(t,\mathbf{r})$.
The Smoluchowski equation for the density of a cluster  is:
\begin{equation}
\frac{d N_k}{dt}=\frac{1}{2}\sum_{i+j=k}K_{i,j}N_i N_j-N_k\sum_{i=1}^{+\infty} K_{i,k}N_i+F(t)\delta_{k,1},
\end{equation}
where $K_{i,j}$ is the reaction rate 
and $\delta_{i,j}$ the Kronecker symbol.
We assume the following  initial conditions: 
if there is no source $F(t)=0$ we have $N_i(t=0)=\delta_{1,i}$, otherwise with a source  we have $N_i(t=0)=0$.
A useful quantity is the moment of order q:
\begin{center}
\begin{math}
M_q(t)=\sum_{i=1}^{+\infty}i^{q}N_i.
\end{math}
\end{center}
The time evolution of the moment is:
\begin{center}
\begin{math}
\dot{M}_q(t)=\frac{1}{2}\sum_{i,j}K_{i,j}((i+j)^q-i^q-j^q)N_iN_j +F(t).
\end{math}
\end{center}
For  $q=1$ the moment  is the total coverage $M_1$,  i.e.  the total
mass, and for $q=0$ it is the total cluster density $M_0$.
 Without source the total mass is conserved 
$M_1(t)=M_1(0)=N_1(0)=1$.
The calculation of the reaction rate depends on some physical assumptions  
The mean-field theory assumes we can
exchange the  many-body problem by a one-body problem with a convenient 
external field. 
 For this purpose we define asymmetric rates,
 the rate that a cluster of mass $i$  aggregates with
 the cluster $j$: $K_{i\rightarrow j}$
(the reaction rate is a symmetric function of $K_{i\rightarrow j}$).
  Following ~[\cite{Bales}], ~[\cite{Pim}]
(we chose a characteristic length instead of a characteristic life-time used in~[\cite{Pim}])
 and dimension analysis, we define the average length of
 a cluster with the following equation:
 \begin{equation}
 \label{kshi}
\frac{\partial N_i}{\partial t}\sim -\frac{D_iN_i}{\xi^2_i}=-
 N_i\sum_{j=0}^{+\infty }K_{i\rightarrow j}N_j,
  \end{equation}
the life-time $\tau_i$ of the cluster is defined with the relation
$\xi_i=\sqrt{D_i\tau_i}$~[\cite{Pim}]. We follow the
mean-field approach ~[\cite{Fried}]; we chose an arbitrary particle $n_i(\mathbf{r},t)$ and 
 the microscopic equation of the cluster reads~[\cite{Bales}]:
  \begin{equation}
  \label{mic}
\frac{N_i-n_i}{\xi^2}=\Delta^{(d)}n_i,
  \end{equation}
 where $\Delta^{(d)}$ is the symmetrical Laplace operator in $d$ dimensions:
\begin{center}
\begin{math}
\Delta^{(d)}=\frac{\partial^2  }{\partial
 r^2}+\frac{d-1}{r}\frac{\partial }{\partial r}.
\end{math}
 \end{center}
The fluctuations 
$(N_i-n_i)$ around the average value $N_i$ are proportional  to the variation
of the microscopic density $\Delta^{(d)}n_i$.
 We consider a spherically symmetric solution of Eq.~(\ref{mic}).
The external field is the same as for the mean-field  approach ~[\cite{Fried}]
, the boundary conditions $n_i(r\rightarrow +\infty)=N_i$ and
 $n(r=r_j)=0$ express the convenient external field.
 Physically it means that close to a cluster the microscopic density
vanishes, i.e we have a perfect sink and far away from a sink the microscopic
density is  close to the average density (we assume an early stage of clustering).
 The stationary spherically symmetric solution of the Eq.~(\ref{mic})
( we  chose as origin the contact  between the cluster $i$ and the point-like cluster $j$) reads~[\cite{Daut}]:
\begin{eqnarray*}
n_i(r)=\left\{ \begin{array}{rcl}
 N_i(1-\exp(-\frac{r-r_j}{\xi_i})) & \mbox{ for }  d=1\\
 N_i(1-\frac{K_0(\frac{r}{\xi_i})}{K_0(\frac{r_j}{\xi_i})})&
 \mbox{ for }  d=2 \\
 N_i(1-\frac{r_j\exp(-\frac{r-r_j}{\xi_i})}{r}) & \mbox{ for }  d=3 \
\end{array}\;\right\}
\end{eqnarray*}
  where $K_0$ is the modified Bessel function ~[\cite{Abr}].
 The reaction rates is proportional to the  flux of cluster $i$ into a 
cluster  $j$ ~[\cite{Fried}] 
$\int_j d\mathbf{r}_j \frac{D_i}{N_i}\frac{\partial n_i}{\partial 
r}|_{r=r_j}$, then:
  \begin{eqnarray*}
\label{sol}
K_{i\rightarrow j}=\left\{ \begin{array}{rcl}
 \frac{D_i}{\xi_i} & \mbox{ for }  d=1\\
\frac{D_i r_j}{\xi_i} \frac{K_1(\frac{r_j}{\xi_i})}{K_0(\frac{r_j}{\xi_i})}
 \sim D_i(1+\frac{r_j}{\xi_i})&
 \mbox{ for }  d=2 \\
 D_i(r_j+\frac{r^2_j}{\xi_i})  & \mbox{ for }  d=3 \
\end{array}\;\right\}
\end{eqnarray*}
 We  have used the asymptotic expansion of the modified Bessel Function
$\frac{xK_1(x)}{K_0(x)}\sim 1+x$ for $x\rightarrow +\infty$ ~[\cite{Abr}]( on length scale of the
order $\xi_i\ll r_j$, i.e.  $x\rightarrow 0$ we have a logarithmic behavior
 $\frac{x K_1(x)}{K_0(x)}\sim -\frac{1}{\ln(x/2)}$ which is typical
 at the critical dimension $d=2$).\\
  The rate for all dimensions is formally:
 \begin{equation}
\label{u}
 K_{i\rightarrow j}=D_i(r_j^{d-2}+\frac{r_j^{d-1}}{\xi_i}).
 \end{equation}
 Eq.~(\ref{u}) contains
  two terms. The first term $r_j^{d-2}$, which is
  (strictly speaking) valid only for $d \geq 3$, we call the mean-field
  term. The second term which is valid only for $d \geq 2$, contains the
   correlation, implicitly defined  in $\xi_i$, we call 
 the correlation term. Eq.~(\ref{u}) is also valid  for $d=1$  if we put $r_j^{d-2}=0$.
One then combines the Eq.~(\ref{u}) with  Eq.~(\ref{kshi}), we get an equation of the second order on $\xi_i$:
\begin{equation}
 \xi_i^2\sum_jr_j^{d-2}N_j+\xi_i\sum_jr_j^{d-1}N_j-1=0.
\end{equation}
 The approximate solution, for a large time scale where the correlation length is short, i.e
$\xi_i\ll r_j$, is\begin{center} $\xi_i=\frac{1}{\sum_jr_j^{d-1}N_j}$.\end{center}
 Notice that the average length $\xi_i$is independent of the mass $i$. It is 
on average the inverse of an effective coverage 
\begin{center} $\sum_jr_j^{d-1}N_j$\end{center} and the life-time 
$\tau_i=\frac{\xi_i^2}{D_i}$ depends on the mass $i$. We obtain  the asymmetric rate:
\begin{center}
\begin{math}
K_{i\rightarrow j}=D_i(r_j^{d-2}+r_j^{d-1}\sum_{k=1}^{+\infty}r_k^{d-1} N_k).
\end{math}
\end{center}
The symmetrization of the reaction rate [\cite{Fried}] leads to:
 \begin{equation}
\label{rate4}
K_{i,j}=(D_i+D_j)((r_j+r_i)^{d-2}+(r_i+r_j)^{d-1}\sum_{k=1}^{+\infty}r_k^{d-1} N_k).
 \end{equation}
One generally assumes that the gyration radius scales with the mass 
$r_i=i^{\alpha}$( in appropriate length units),
where $\alpha=1/d_f$ is the inverse of the fractal dimension of the boundary of the cluster(
 for compact island $d_f=d-1$, and
 $d_f=1.72$ for Diffusion Limited Aggregation process in in $d=2$~[\cite{Bales}]).
The effective coverage is for compact islands, $M_{(d-1)/d_f}$ is close to the
coverage $M_1$ when $d\gg 1$.\\
From the Eq.~(\ref{rate4}) we define two rates: the  mean-field rate 
$K_{i,j}^{MF}=(D_i+D_j)(r_j+r_i)^{d-2}$
and the  rate with correlation  $K_{i,j}^{C}=(D_i+D_j)(r_j+r_i)^{d-1}M_{\alpha(d-1)}$
. The difference between these both rates is that we exchange  a spatial
 dimension $(r_j+r_i)$ in mean-field rate through a moment $M_{\alpha(d-1)}$ 
 in the rate with correlation.
We have hence a Taylor expansion around the correlation number $\frac{M_{\alpha(d-1)}}{\xi_i}$.
In $d=1$ only the correlation term remains and the rate is:
 \begin{equation}
\label{rate3}
K_{i,j}=(D_i+D_j)M_o.
 \end{equation}
We assume $M_{\alpha(d-1)}\rightarrow M_0$ when $d\rightarrow 1$, because the scaling effect of the radius of gyration  vanishes in $d=1$, i.e.  the boundary for an 
aggregating particle is point-like.\\
For a time scale larger than $t_c$
, defined by $M_{\alpha(d-1)}(t_c)\sim(r_i+r_j)$, the correlations dominate,
hence the reaction rate is essentially:
\begin{equation}
\label{rate5}
K_{i,j}=K_{i,j}^{C}=(D_i+D_j)(r_j+r_i)^{d-1}M_{\alpha(d-1)}.
\end{equation}
We can derive solutions of the Smoluchowski equation with only rate with correlation Eq.~(\ref{rate5})
from known solutions with mean-field rate $K_{i,j}^{MF}=(D_i+D_j)(r_j+r_i)^{d-2}$ in a  special case,
without source ($F=0$) and with a rate given from Eq.~(\ref{rate5})the 
solution of the Smoluchowski equation in $d$ dimensions is equivalent 
to the solution with the mean-field rate $K^{MF}$ without source ($F=0$),
up to a time rescaling $d \tilde{t} =M_{\alpha(d-1)}(t)dt$, in $d+1$ dimensions.\\

%
%
%
%
With our method a large class of processes in $d=1$ can be solved analytically.\\
In order to confirm this approach we present some analytical solutions and
numerical evidence.

\begin{center}
 \begin{enumerate}
 \item
The solution of the Smoluchowski equation with $K_{i,j}=1$ and without source ($F=0$) is known analytically ~[\cite{Fried}],  $N_k\sim t^{-2}$ and $M_0 \sim t^{-1}$,
 hence the  time rescaling $d\tilde{t}=M_0(t)dt$ leads us to the known 
solution in $d=1$ of the clustering process with mass-independent diffusion.
  This solution was found with another method  ~[\cite{Hin}] is
 $M_0 \sim t^{-1/2}$
\begin{figure}
  \includegraphics[width=8.2cm,angle=-90]{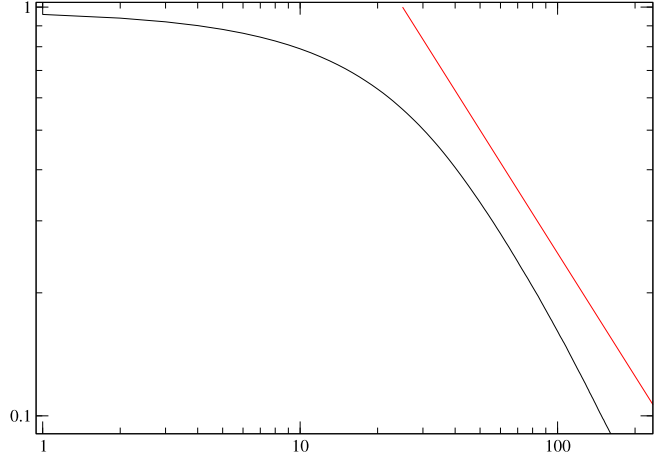}
 \caption
{\footnotesize  \label{fig1}
{\small  Cluster density $N_1$ as a function 
of time (for comparison the red curve is $\sim t^{-1}$)}}
\end{figure}. 
Notice that 
through our method we obtain the asymptotic behavior for all clusters, $N_k\sim t^{-1}$
(~\figurename{\ref{fig1}})
which was not possible with the method used in ~[\cite{Hin}].

\item With constant source $F=const$, and $D_i=1$ we have the following equations:
\begin{center}
\begin{math}
\dot{N}_k=M_0\sum_{i+j=k}N_iN_j-2N_kM^2_0+\delta_{1,k}F
\mbox{ and }
\dot{M}_0=-M^3_0+F.
\end{math}
\end{center}
The asymptotic solutions are $M_0\sim F^{\frac{1}{3}}$ 
(~\figurename{\ref{a1}}) and 
$N_k\sim k^{-\frac{3}{2}}$ (~\figurename{\ref{b1}})
. For comparison the solution with the mean-field 
term ~[\cite{Kra1,Kra2}]in $d=2$ is $N\sim F^{\frac{1}{2}}$  and $N_k \sim k^{-3/2}$.
\begin{figure}
  \includegraphics[width=8.2cm,angle=0]{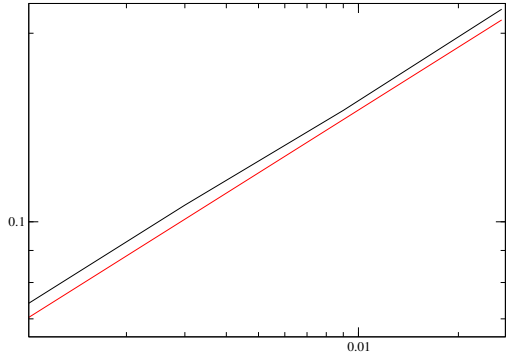}
 \caption
{\footnotesize 
 \label{a1}{\small  Asymptotic value of the total cluster density $M_0$ as a function of the
flux $F$ (for comparison the red line has a slope $F^{1/3}$)}}
\end{figure}
\begin{figure}
  \includegraphics[width=8.2cm,angle=0]{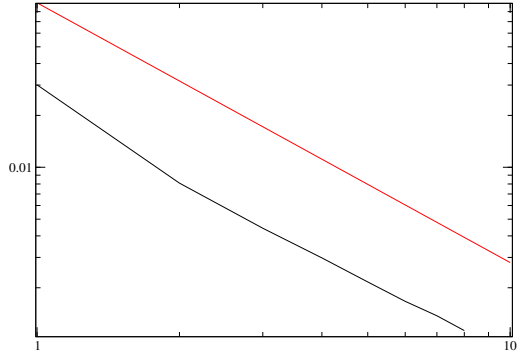}
 \caption
{\footnotesize  \label{b1}
{\small   Asymptotic value of the cluster density $N_k$ as function 
of the cluster mass $k$(for comparison the straight line $\sim k^{-3/2}$)}}
\end{figure}
\item
For the sake of completeness we present an analytical solution already known: $D_k=\delta_{1,k}$ 
and source $F=const$. The solutions were  found with another method in ~[\cite{Pim}] 
are $N_1\sim t^{-1/2}$ and total cluster density $M_0 \sim t^{1/4}$. With our method we obtain 
the same rates and therefore the same solution. For comparison the mean-field
approach gives $N_1\sim t^{-1/3}$ and $M_0\sim t^{1/3}$ which are different to
the numerical solution and the experimental result.
 \end{enumerate}\end{center}
Hence we have shown that this approach  leads to the correct description
of irreversible aggregation with brownian diffusion in one dimension
for different models. 
We hope to find similar results in $d>1$ beyond the time 
scale where the mean-field rate is valid, i.e. for $t>>t_c$.\\
 There are some open questions:
\begin{center}
 \begin{enumerate}
 \item {\it Higher order of the Taylor expansion:}
can we compute other rates when the correlation between cluster 
is more important, i.e. can we obtain the
 full Taylor expansion around the correlation parameter
 $\frac{M_{\alpha(d-1)}}{\xi_i}$
Perhaps larger correlation effect can be obtained by 
modifying the boundary condition of Eq.~(\ref{mic})
 \item {\it multiple particles reaction: } how can this
 approach  be generalized to 
systems where the binding energy is so tight that binary reaction
is negligible and we have to consider the
 reaction  $A_1+\cdots+A_n\rightarrow B_1+\cdots+B_m$
~[\cite{Na}] with an arbitrary $n$ and $m$ ?
 \end{enumerate}\end{center}

\end{document}